# Surface plasmon assisted magnetic anomalies on room temperature gold films in high-intensity laser fields

N. Kroó, P. Rácz AND S. Varró

*Wigner Research Centre for Physics, Konkoly-Thege M. út 29-33, 1121 Budapest, Hungary*

**Abstract**

Supplementing our STM and electron emission studies investigations, concluding in electron pairing in strong laser fields [1], further time-of-flight electron emission studies were carried out, changing the angle of polarization of the incident light, exciting surface plasmon oscillations. It has been found, that those parts of the electron spectrum which have been attributed to electron pairing have a significantly different angular dependence around 80 GW/cm$^2$ where the pairing effect has been found than outside this region (e.g. 120 GW/cm$^2$). These results have been interpreted as the appearance of ideal or partly ideal diamagnetism on the one hand and as anomaly in the magneto-optical effect (rotation) on the other, in the same laser intensity region where the pairing effect has been found.

In recent years the number of papers on the study of surface plasmon oscillations (SPO) has considerably increased. This phenomenon is connected partly with the unique properties of these collective excitations of conduction electrons on the surface of some metals, coupled to an evanescent electromagnetic field above the surface. These properties open up a broad spectrum of applications in information and communication technologies, chemistry, biology, medicine, etc. [2,3]. But in addition, due to the fact that SPO-s enhance the electromagnetic field impinging onto to the surface, a goldmine of different nonlinear effects sets in at much lower laser intensities than in nonplasmonic cases and can be studied with several experimental techniques.

In a recent paper [1] we reported on one such effect, namely an indication to surface plasmon assisted electron pairing in gold films at room temperature. The SPO-s were excited by the p-polarized light of femtosecond Ti:Sa lasers in the Kretschmann geometry, and STM measurements were carried out at near field hot spots on the gold surface where we registered the rectified near field at zero STM bias [4]. Additional measurements were also carried out, studying the SPO induced electron emission from the same gold surface with the time-of–flight technique.

It was found in the STM measurements that the response signal of the microscope, when changing the intensity of the 120 fs laser pulses increased from ~15 microsecond to above 60 microsecond, peaking around 70-80 GW/cm$^2$ laser intensity. A decreased probability of SPO excitation was also found in the same laser intensity range. Furthermore, a separate peak was found in the time-of–flight spectrum of surface plasmon assisted electron emission, which may be attributed to the same pairing phenomenon. The intensity of this peak changed more than two orders of magnitude around 70-80 GW/cm$^2$ laser intensity.

These experimental findings were interpreted on the basis of an analogy with an earlier theoretical result on e-e scattering in strong laser field in vacuum [5]. In this theory an effective potential has been derived for each multiphoton absorption (or stimulated emission) channels, containing a Bessel function factor with negative regions, implying electron attraction in these negative parts of the relative potential energy of the electrons [1, 5]. Concerning the intensity range, where the anomalous behavior of both the STM and TOF signals were observed, a reasonable agreement was found with our experimental data if an SPO field enhancement factor of 30 was used (which has been taken from one of our previous papers [6]). In addition to this mechanism of electron pairing there might potentially be another one, namely that of pairing by the spin fluctuation mechanism [7,8]. The interpretation of our observations, however, need not necessarily rely on this possibility since this latter mechanism is connected with the pairing of near neighboring spins, while in our case the experimental data are in agreement with the theoretically deduced effective electron-electron attraction at a significantly larger distance. In order to find further details in connection with the observed phenomenon, additional time-of-flight measurements were carried out, varying the plane of polarization of the incident laser light.

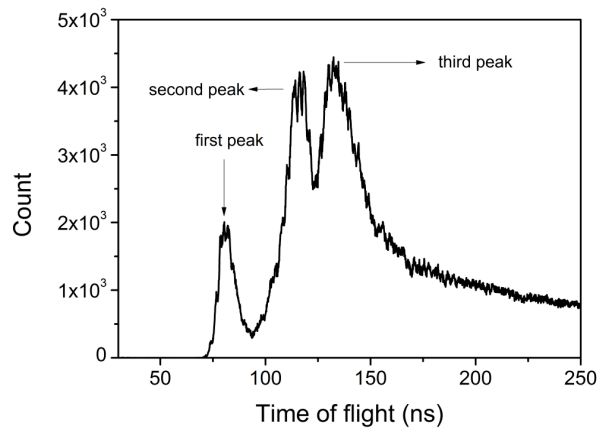

**Fig. 1**: Typical measured time-of-flight spectrum of electrons emitted from a gold metal film evaporated on the top of a right angle prism (Kretschmann configuration). Three separated peaks can be observed in the spectrum. Laser intensity was 80 GW/cm$^2$, and pulse length was 40 fs. The total number of electrons emitted into the first (high energy) peak is related to the electron pairing effect.

In the present paper we report on our results of these measurements. The surface plasmon induced electron emission experiments had been studied by exciting the SPO-s by Ti:Sa based chirped-pulse amplifier-laser systems. The amplifiers deliver 40 fs pulses with 1 kHz [1] and 30 fs pulses with 5 kHz [9] at 800 nm central wavelength. The pulse energy was reduced to below 1 μJ. The SPO-s were generated in the Kretschmann geometry and for this purpose a right-angle fused silica prism was used with 45nm gold film evaporated on the hypotenuse surface. The prism was used as the window of a vacuum chamber with the gold plated surface inside the chamber (the pressure in the chamber was around 10$^{-7}$ mbar). The intense laser beam excitation resulted in SPO mediated electron emission. The electrons, emitted were measured by a time-of-flight electron spectrometer (Kaesdorf Electron TOF) in this vacuum chamber. Fig 3(a) illustrates this experimental setup. Due to the relatively large surface area where the plasmons were excited, our measurements represent the average of emission from numerous hot spots and from other parts of the surface with much lower SPO field-enhancement. The acceptance cone of the spectrometer was relatively high with an accelerating voltage applied inside the TOF spectrometer. Therefore it was possible to measure the total photocurrent as a function of the incident laser intensity as well as the spectrum of electrons simultaneously. One typical measurement of the time-of-flight spectrum of emitted electrons is shown in Fig. 1 where 3 peaks have been found.

The first, highest energy, smallest intensity one has been attributed to electron pairing [1]. The qualitative explanation of the higher energy of these electrons is, that they are emitted from the vicinity of the Fermi-surface, where the density of state may have a maximum [10]. The area of this high energy peak as a function of laser intensity could be explained on the basis of the analogy with an attractive potential, describing electron-electron scattering in high-intensity radiation fields [1,5]. This peak represents only a few percent of the total emitted electron current. The observation of higher electron energies had earlier been reported in several papers [6,11,12,13,14] where the emission mechanisms had also been discussed. Our conclusions are, however, independent of the emission (and acceleration) mechanism. Since no electron emission has been observed with s-polarized light excitation, this is a strong argument against the potential thermal origin of our finding, similarly as the exciting laser intensity dependent peaking of this part of the TOF spectrum around 70-80 GW/cm$^2$ laser intensity. As far as the work function of gold is concerned even the lowest value is above 4.6eV, but the typical one is around 5.1 eV. In both cases, however, four 1.5 eV photons and (so SPO-s) are needed to get an electron out of the metal surface [15]. In agreement with this expectation in the intensity range, where the perturbation theory may be applied, the "n-power law" dependence on the laser intensity of the total electron number has been found with a slope of ~4.3, if plotted on double logarithmic scale (Fig 2 (a)). At higher laser intensities, however, the emission mechanism turns into the tunneling regime, similarly as in earlier findings [16] but this change sets in at much lower incoming laser intensities than e.g. in gases. This decrease is attributed to the field enhancement effect of surface plasmons. The analysis of the middle peak of the TOF spectrum will be carried out elsewhere. We mention here only, that both the low and middle energy peaks show the same character as the total intensity, turning into the tunneling regime around 80 GW/cm$^2$ and do not show the peaking property like the high energy peak shown in Fig2(b).

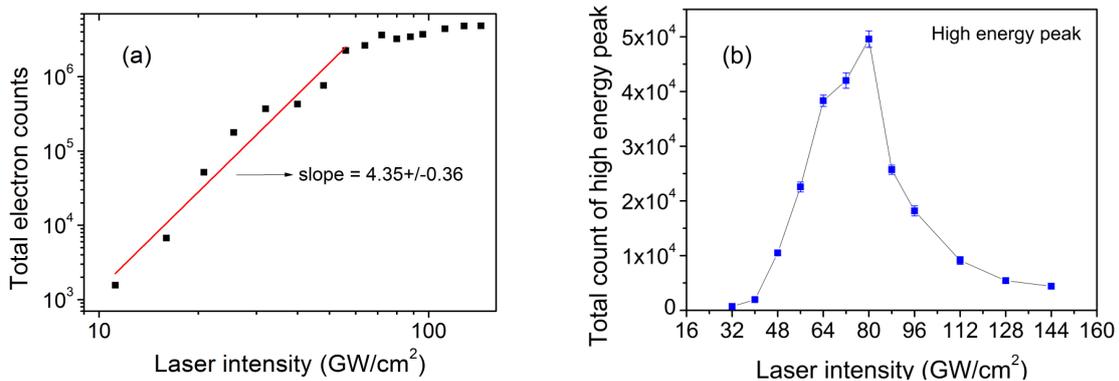

**Fig. 2**: (a): The laser intensity dependence of the total number of emitted electrons, and (b): the total number of electrons, emitted into the higher energy peak (first peak in Fig. 1) of the electron spectra recorded in p polarization light (zero angle).

As stated above, in order to get further information, time-of-flight measurements were carried out, varying the plane of polarization of the incident light. When the electric field vector of the incoming light is perpendicular to the gold film, the magnetic field is parallel with the surface and has no effect on SPO-s. When turning the polarization plane toward s-polarization the situation changes. In these cases of polarization the projection of the k-vector of the incoming light, along the propagation vector of the SPO does not change, therefore the SPO resonance condition still remains fulfilled. There is also an electric field component in the target surface plane, perpendicular to the plane of incidence, and it has also no effect on the SPO generation. A magnetic field, vertical to the surface, is also emerging at these nonzero polarization angles of the electric field, and it increases with increasing angles. And since in the used laser intensity range the EM field gets rectified, we may assume that a dc magnetic component is also present. This field may influence the SPO mediated electron emission due to the influence of the SPO parameters [17]. We may expect e.g. a transition of the gold sample towards a diamagnetic state. A magneto-optical rotation effect may also be expected. This latter may lead to an asymmetry in the polarization angle-dependence of the electron emission. We emphasize this since it is generally believed, that gold is already without any external influence an ideal diamagnetic material and therefore this expected effect should not exist. It has been found recently, however, that bulk gold shows paramagnetic properties too [18].This is even especially true for nanostructures, which show strong magnetism [19]. This was just our motivation for performing light polarization-dependence experiments in the laser intensity range around 80GW/cm$^2$.

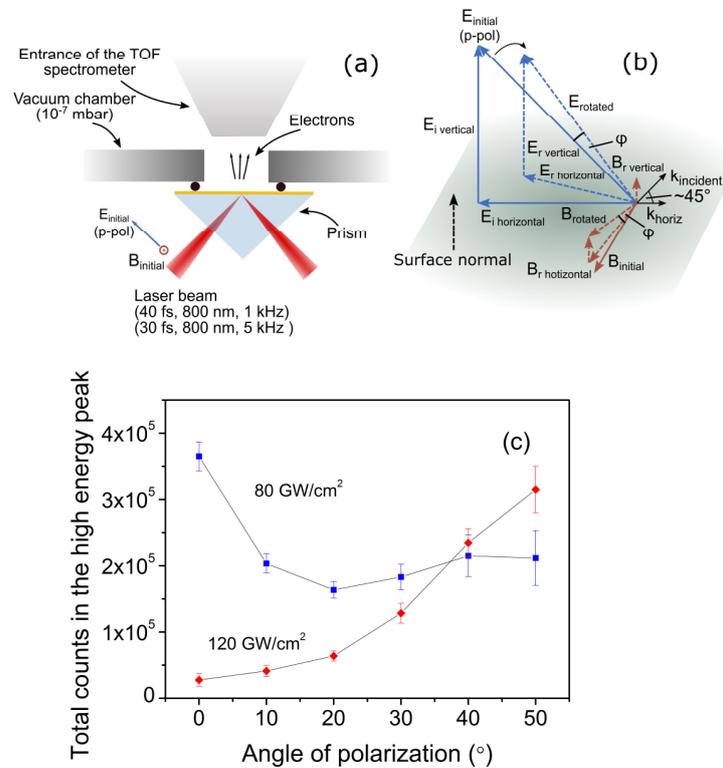

**Fig. 3**: (a): The layout of the experiment. (b): The electric and magnetic field directions at zero (p-polarization) and rotated directions. (c): The polarization dependence of the total counts of the high energy peaks at different laser intensities. The data at 80 GW/cm$^2$ intensity are compared with those at 120 GW/cm$^2$. These results are in agreement with the expected expelled magnetic field. See further details in the text.

Now let us see if our experimental data support these expectations? One of the typical results of the laser intensity dependence at p-polarization of the observed higher energy peak in the TOF spectrum is shown in Fig 2(b), and the polarization dependence measurements were carried out in this laser intensity region. It is known, that magnetic fields may influence the surface plasmon properties [20,21]. When the plane of polarization of our laser light is perpendicular to the gold surface (0 degree, i.e. p-polarization), the magnetic field is in the gold surface plane and has no influence on the SPO properties. If we turn, however, the plane of polarization around the momentum vector of the propagating SPO-s, a vertical magnetic component is created which perturbates strongly these SPO-s. Fig.3 shows, however, that when the vertical component of the electric field at 120 GW/cm$^2$ is equal to that of the 80 GW/cm$^2$ and p-polarization one, the area of the high energy peak is equal in the two cases within experimental errors. Several of our similar observations prove, that the magnetic field has no effect on this peak i.e. this field has seemingly been expelled from the film at 80 GW/cm$^2$.

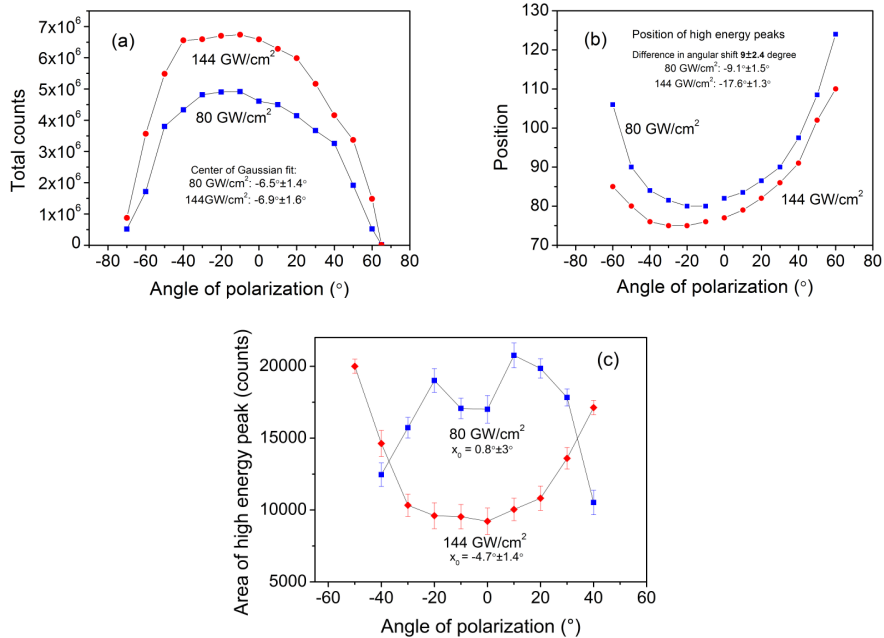

**Fig. 4**: (a): The total number of electrons and (b): the position (time of flight in nanoseconds displayed on the abscissas of Fig. 1) of high energy peaks in the time of flight spectra as a function of the angle of polarization (in clockwise and anticlockwise directions) at different laser intensities (blue squares at 80 GW/cm$^2$ and red dots at 144 GW/cm$^2$ intensity of the 40fs laser). (c): polarization angle dependence of the intensity of the high energy peaks as measured with the 30 fs laser. The angular shift is smaller here at 144 GW/cm$^2$ than in Fig. 4(b), but still significant. The shift at 80 GW/cm$^2$ is practically zero. Based on the asymmetry of these angular dependence measurements magneto-optic rotation anomaly may be concluded around 80 GW/cm$^2$. See further details in the text.

In some further TOF measurements we looked for the polarization angle dependence of the observed spectra, by changing the polarization plane of the incident laser light both in clockwise and anticlockwise directions. As shown in Fig 4(a), the distribution of the total intensity is nearly symmetric, centered around ~6.5° (the prism with the gold film has not been perfectly aligned to zero degree). The position of the high energy peak, however, is shifted about 11.5$^0$ relative to this zero level in the anticlockwise direction at 144 GW/cm$^2$ compared to that at 80 GW/cm$^2$, which has been shifted only by a much lower amount, namely about 2.5$^0$. This is shown in Fig 4(b) where Gaussian curves were fitted to both data. About the same shift, namely 8.5° is observed in the difference of the polarization dependence of the peak intensities of the two cases. Nearly the same shift difference, namely 5.5$^0$ has been found in the polarization angle dependence of the area of the high energy peaks (Fig 4c) with a 30 fs laser. And here the 144G W/cm$^2$ data around 40 degree reach also the level of the 0 degree 80 GW/cm$^2$ data, indicating again, as in Fig.3, that the vertical to the gold surface magnetic field does not influence the results, i.e. it is thought to be expelled from the film.

On the basis of the described observations the following conclusions may be drawn: In addition to the conclusion on electron pairing of our paper published recently [1], two additional statements can be made. From the polarization dependence of the high energy peaks in our TOF spectra we might state, that the magnetic field is expelled from the gold film around 80 GW/cm$^2$ laser intensity, i.e. the film is transformed into diamagnetic phase, i.e. we found an indication to the appearance of the Meissner effect in the laser intensity range around 80 GW/cm$^2$.

From the asymmetry of the angular dependence data, as measured at positive and negative angles of the polarization plane of the incident laser light, we may conclude, that a significant magneto-optic rotation exists at 144 GW/cm$^2$ compared to that at 80 GW/cm$^2$, indicating again the same anomalous influence of the magnetic field.

The experimentally found 3 effects are characteristic features of materials being in the superconducting state and we do not know any other phenomenon with the same properties. This is why our potential conclusion may be, that the intense laser fields in the given intensity range turn the gold film into the superconductive state at room temperature.

**Acknowledgements:**

The authors are thankful to Prof. Andrius Baltuška for offering one of his laser at TU Vienna for control measurements of the observed magnetic anomalies (Meissner effect and magneto-optical polarization rotation). Sándor Varró and Péter Rácz thank the support by the Hungarian Scientific Research Foundation OTKA, (Grant No. K 104260 and PD 109472)